\documentclass[times,final]{elsarticle}
\usepackage{jcomp}
\usepackage{framed,multirow}

\usepackage{latexsym}
\usepackage{url}
\usepackage{algorithm}% http://ctan.org/pkg/algorithms
\usepackage{algpseudocode}% http://ctan.org/pkg/algorithmicx
\usepackage{amsmath,amssymb,amsfonts}
\usepackage{graphicx}
\usepackage{textcomp}
\usepackage{xcolor}
\usepackage{booktabs} 
\usepackage{amsmath}
\usepackage{graphicx}
\usepackage{subfigure}
\usepackage{multirow}
\usepackage{amsfonts,bm}
\usepackage{braket}
\usepackage{mathtools}
\usepackage{enumerate}
\usepackage{enumitem}
\usepackage{empheq}
\usepackage{calc}
\usepackage{balance}
\usepackage{dsfont}
\usepackage{textcomp} 
\usepackage{caption}
\usepackage{color, colortbl}
\usepackage{adjustbox}
\usepackage{url}
\usepackage[utf8]{inputenc}
\definecolor{newcolor}{rgb}{.8,.349,.1}

\journal{Journal of Computational Physics}

\begin{document}

\verso{Given-name Surname \textit{etal}}

\begin{frontmatter}

\title{\Large HOSSnet: an Efficient Physics-Guided Neural Network for Simulating Crack Propagation }%
\tnotetext[tnote1]{This is an example for title footnote coding.}

\author[1,2]{Shengyu Chen\corref{cor1}}
\cortext[cor1]{Corresponding author}
\author[1]{Shihang Feng\corref{cor1}}
\author[1]{Yao Huang}
%% Third author's email
\author[1]{Zhou Lei}
\author[2]{Xiaowei Jia}
\author[1]{Youzuo Lin}

\author[1]{Estaben Rougier}

\address[1]{Earth and Environmental Science Division, Los Alamos National Laboratory, Los Alamos, NM  87545, USA}
\address[2]{Department of Computer Science, University of Pittsburgh, Pittsburgh, PA, 15260}

\received{1 May 2023}
\finalform{10 May 2023}
\accepted{13 May 2023}
\availableonline{15 May 2023}
\communicated{XXX}

\begin{abstract}
Hybrid Optimization Software Suite (HOSS), which is a combined finite-discrete element method (FDEM), is one of the advanced approaches to simulating high-fidelity fracture and fragmentation processes but the application of pure HOSS simulation is computationally expensive. At the same time, machine learning methods, shown tremendous success in several scientific problems, are increasingly being considered promising alternatives to physics-based models in the scientific domains. Thus, our goal in this work is to build a new data-driven methodology to reconstruct the crack fracture accurately in the spatial and temporal fields. We leverage physical constraints to regularize the fracture propagation in the long-term reconstruction. In addition, we introduce perceptual loss and several extra pure machine learning optimization approaches to improve the reconstruction performance of fracture data further.  We demonstrate the effectiveness of our proposed method through both extrapolation and interpolation experiments. The results confirm that our proposed method can reconstruct high-fidelity fracture data over space and time in terms of pixel-wise reconstruction error and structural similarity. Visual comparisons also show promising results in long-term reconstruction. 

\end{abstract}

%\begin{keyword}
%% Keywords
%\KWD Keyword1\sep Keyword2\sep Keyword3
%\end{keyword}

\end{frontmatter}

%\linenumbers


\section{Introduction}
Knowledge of brittle materials failure is significant since they have been widely applied in a variety of areas, including structural engineering, geotechnical engineering, mechanical engineering, geothermal engineering, and the oil \& gas industry. The mechanism of brittle material failure is strongly affected by the microstructure of the material, especially the pre-existing micro cracks~\cite{bazant2019fracture,petersson1981crack}. The dynamic propagation and interaction behavior of micro-cracks is essential to estimate the failure of brittle material~\cite{brooks2013fracture,vesely2007structural}. The growth and coalescence of micro-cracks will result in a complex stress state near the crack tips, leading to the catastrophic failure of brittle material in macro-scale~\cite{freiman2019fracture,lamon2016brittle}.

Many approaches have been proposed in the literature to analyze the crack initiation, and propagation behavior in brittle materials, including theoretical constitutive models and numerical methods~\cite{rice1968mathematical,rice1974ductile,hutchinson1968plastic,hutchinson1968singular,hutchinson1991mixed,xu1994numerical,li1985comparison,buehler2003hyperelasticity,ravi1997role,desroches1994crack}. The finite-discrete element method (FDEM) is one of the advanced approaches~\cite{hillerborg1976analysis}. Based on the FDEM, the Hybrid Optimization Software Suite (HOSS)~\cite{knight2016user} is developed as a fracture simulator. This high-fidelity model simulates fracture and fragmentation processes or materials deformation in both 2D and 3D complex systems, providing accurate predictions of fracture growth and material failure. However, the application of pure HOSS simulation has computational limitations since they resolve all the individual cracks with highly resolved meshes and small time steps at large scales. Hence, a new approach can provide key quantities of interest with more efficiency but retain reasonable accuracy in demand.

Machine learning (ML) models are gaining significant attention as a promising alternative to physical simulations in recent years due to their remarkable success and high efficiency in various applications, such as image segmentation~\cite{badrinarayanan2017segnet, chen2018encoder, yang2020improved} and image reconstruction~\cite{haggstrom2019deeppet, montalt2021machine, lin2021artificial}. Encoder-decoder networks~\cite{read2019process} have been employed to simulate the intricate two-dimensional subsurface fluid dynamics occurring in porous media~\cite{thavarajah2021fast}. Fourier Neural Operator~\cite{li2020fourier}, which combines Fourier analysis and neural networks, has been utilized to solve complex physical problems, such as wave propagation~\cite{li2022solving} and multiphase flow dynamics~\cite{wen2022u}. Temporal models, such as the long short-term memory (LSTM) model, have been extensively utilized to capture long-term dependencies in temporal propagation. For instance, in the hydrology domain, the LSTM model has been broadly used to model temporal patterns of water dynamics~\cite{kratzert2019toward,jia2019sdm2}.

%Physical informed neural network
%~\cite{rasht2022physics}  ~\cite{cai2021physics} 
%Navier–Stokes formulations for incompressible flow

To speed up the simulation of HOSS, ML approaches had been utilized to emulate the high-ﬁdelity model efﬁciently. The micro-cracks are represented by a feature vector, then a variety of ML methods, such as decision trees, random forest, and artificial neural networks, are utilized to predict the time and the location of fracture coalescence~\cite{moore2018predictive,srinivasan2018quantifying,hunter2019reduced}. But the fracture damage, which is the key input needed for upscaling micro-scale information to macro-scale models, is not estimated. A graph convolutional network (GCNs), coupled with a recurrent neural network (RNN), was used to model the evolution of those features from the reduced graph representation of large fracture networks~\cite{schwarzer2019learning}, where the sum of the fracture damage in each finite element mesh is represented as total damage.
However, these works only consider using the fracture features to predict its propagation.
%surrogate models~\cite{mudunuru2019surrogate}

In this paper, we develop a new data-driven method, termed Hybrid Optimization Software Suite Network (HOSSnet), to improve the high-fidelity of micro-crack fracture reconstruction from Cauchy stress and fracture damage in spatial and temporal fields. We also leverage the underlying physical constraints to further regularize the model learning of generalizable spatial and temporal patterns in the reconstruction process. In particular, our proposed method consists of two components: high-fidelity reconstruction unit (HRU) and physics-guided regularization. HRU is designed based on an Unet-based~\cite{barkau1996unet} encoder-decoder structure to improve the reconstruction performance in the spatial field. HRU also uses an LSTM layer to capture long-term temporal dependencies. In addition, The physics-guided regularization method adjusts the reconstructed data over time by enforcing consistency with known physical constraints such as optical flow~\cite{feng2021connect} and positive direction.

Our evaluations of the HOSS dataset~\cite{knight2016user} have shown the promising performance of the proposed HOSSnet over space and time in both interpolation and extrapolation experiments. We also demonstrate the effectiveness of each component of our design by showing the improvement both qualitatively and quantitatively.

\section{Related Work}
%\subsection{Overview of Micro-crack problem}

%HOSS~\cite{knight2016user} is a fracture simulator generated based on the combined FDEM to model fragmentation in brittle materials, especially with the pre-existed micro cracks. HOSS simulations have been applied to analyze fracture initiation and propagation behavior in shale rocks in order to enhance the efficiency of hydraulic fracturing~\cite{carey2014shale} and to model earthquake ruptures~\cite{okubo2017modelling}. The FDEM formulation describes fracture and fragmentation processes explicitly based on conservation laws, avoiding unnecessary assumptions regarding the behavior of the model.

%In the previous works, different approaches have been proposed to  approaches have been applied to emulate the high-fidelity model efficiently, including machine learning~\cite{moore2018predictive,schwarzer2019learning}, statistics-based constitutive model~\cite{vaughn2019statistically}, and Graph Theory~\cite{mudunuru2018estimating}. The key problems~\cite{moore2018predictive} focused on were time to failure and fracture coalescence predictions. The previous work~\cite{moore2018predictive} did not estimate damage, which is a key input needed for upscaling micro-scale information to macro-scale models.

\subsection{Machine Learning Approaches in Scientific Domain}

Machine learning (ML) models,  given their tremendous success in several commercial applications (e.g., computer vision, natural language processing, etc.), are increasingly being considered promising alternatives to physics-based models in the scientific domains. For example, in the hydrology domain, researchers have used graph neural networks (GNN) in modeling spatial dependencies~\cite{moshe2020hydronets,chen2021heterogeneous} of river networks. In streamflow problems, Moshe et al.~\cite{moshe2020hydronets} proposed the HydroNets model, which uses ML models to integrate the information from river segments and their upstream segments to improve the streamflow predictions. Chen et al.~\cite{chen2021heterogeneous} proposed a Heterogeneous Stream-reservoir Graph Network (HRGN) model to represent underlying stream-reservoir networks and improve streamflow temperature prediction in all river segments within a river network. 

The convolutional neural network (CNN) based and generative adversarial networks (GAN) based models have been widely used in high-resolution turbulent flow simulation. Fukami et al.~\cite{fukami2019super} propose an improved CNN-based hybrid downsampled skip-connection/multi-scale (DSC/MS) model by extracting patterns from multiple scales. This method has been shown to produce a good performance in reconstructing the turbulent velocity and vorticity fields from extremely low-resolution input data. Similarly, Liu et al.~\cite{liu2020deep} also propose another CNN-based multiple temporal paths convolutional neural network (MTPC) to simultaneously handle spatial and temporal information in turbulent flow simultaneously to fully capture features in different time ranges. Deng et al.~\cite{Deng2019SuperresolutionRO} demonstrate that both super-resolution generative adversarial networks (SRGAN) and enhanced super-resolution generative adversarial networks (ESRGAN)~\cite{wang2018esrgan} can 
produce a reasonable reconstruction of high-resolution turbulent flow in their datasets.

In the micro-crack problem,  Perera~\cite{perera2022graph} proposed a graph neural network to represent the relationship of different cracks and initially realize simulating cracks propagation for a wide range of initial microcrack configurations. This method only considers the cracks propagation from fracture features to fracture features but cannot apply to the model mapping from Cauchy features to fracture features. %Compared with this method, Our proposed HOSSnet method can achieve both model mapping accurately and get promising performance.

\subsection{Physics-based Loss Function}

There are still several challenges faced by existing machine learning methods. Standard machine learning models can fail to capture complex relationships amongst physical variables. Results will be even worse if only contain limited observation data. This is one reason for their failure to generalize to scenarios not encountered in training data. Hence, researchers are beginning to incorporate physical knowledge into loss functions to help machine learning models capture generalizable dynamic patterns consistent with established physical relationships. In recent studies~\cite{willard2020integrating,cuomo2022scientific}, the use of physical-based loss functions has already shown promising results in a variety of scientific disciplines. For example, Karpatne et al.~\cite{karpatne2017physics} propose an additional physics-based penalty based on known monotonic physical  relationship to guarantee that the density of water at lower depth is always greater than the density of water in any depths above. Kahana et al.~\cite{Adar2020physical} apply an additional loss function to ensure the physical consistency in the time evolution of waves, improving the prediction results and making the model more robust. %Inspired by this idea,  we introduce the physical constraints (optical flow~\cite{feng2021connect} and positive direction) to regularize the fracture reconstruction in the long-term propagation. 

\section{Problem Definition}

The objective of our proposed work is to reconstruct the missing fracture image $\textbf{Y}$ for each micro-crack $m\in \{1,..., M\}$ at each time step $t\in \{1,..., T\}$, given input physical variables that drive the dynamics of the physical system. In detail, we use $\textbf{X}_c = \{\textbf{x}_{m,c}^t\}$ to represent dynamic Cauchy input features for each micro-crack $m$ at a specific time step $t$. $\textbf{X}_c$ includes three Cauchy features of all micro-crack: Cauchy1 $C_x$, Cauchy2 $C_y$, and Cauchy12 $C_{xy}$ %, representing cauchy vectors in $x$, $y$ and $xy$ direction respectively. 
In another case, we also regard $\textbf{X}_f = \{\textbf{x}_{m,f}^t\}$ as a dynamic fracture input feature for each micro-crack on each time step. Then we use these two different groups of input features: $\textbf{X}_c$ and $\textbf{X}_f$ to reconstruct the missing target variables (i.e., fracture) $\textbf{Y}  = \{y_m^t\}$ for certain micro-crack on certain time step respectively.  When $\textbf{X}_c$ is used as the input, we represent this scenario as Cauchy $\rightarrow$ Fracture. On the contrary, the scenario would be Fracture $\rightarrow$ Fracture if $\textbf{X}_f$ is the input.

\begin{figure*}[htbp]
\begin{center}
% \raggedleft
\includegraphics[width = 1.0\linewidth]{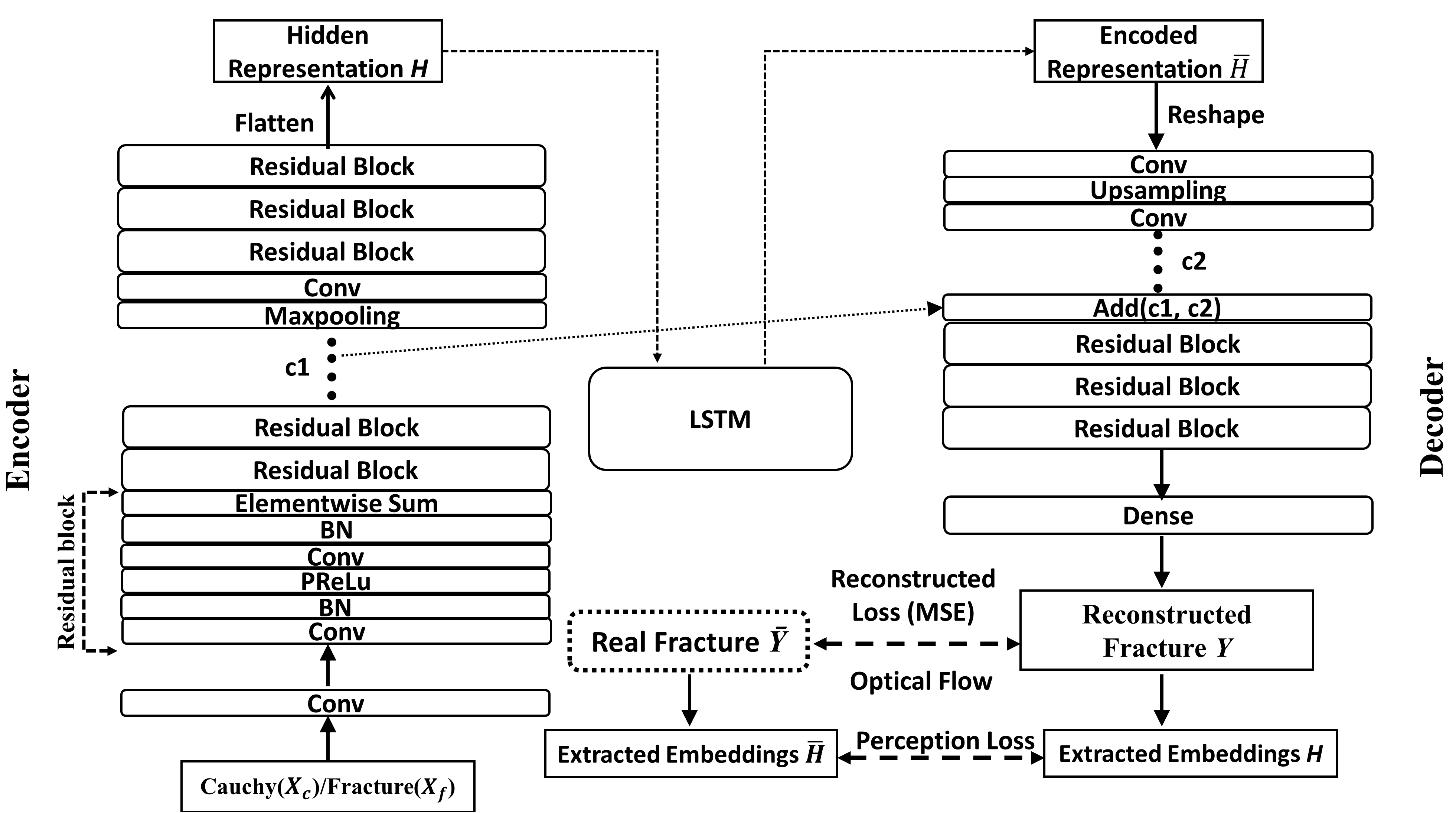}
\end{center}
\caption{The architecture of the proposed HOSSnet model and different components in the loss function.  }
\label{fig:tf_plot}
\end{figure*}

\section{Methods}
The purpose of our proposed HOSSnet model is used to achieve an end-to-end reconstruction  mechanism from input feature $\textbf{X}$ (Cauchy feature $\textbf{X}_c$ or fracture feature $\textbf{X}_f$) to missing target variable (i.e., fracture) $\textbf{Y}$ over the time and over the sample. In Figure.\ref{fig:tf_plot}. we show the overall structure of the proposed HOSSnet method.  These are described here, in order:

\subsection{HOSSnet Architecture}
This proposed HOSSnet aims to learn the model mapping from $\textbf{X}$ to $\textbf{Y}$, containing two main components: HRU and physics-guided regularization. 

HRU is designed based on benchmark Unet-based ~\cite{barkau1996unet} encoder-decoder structure. It contains a contracting path (encoder), recurrent network layer (RTL), and expansive path (decoder). The contracting path consists of the repeated application of two 3x3 convolutions (unpadded convolutions), each followed by a rectified linear unit (ReLU) ~\cite{enwiki:1132964988} and a 2x2 max pooling operation with stride 2 for downsampling. Additionally, we also introduce multiple residual blocks between every two convolutions. And each residual block contains convolutional layers~\cite{o2015introduction}, batch normalization layers~\cite{ioffe2015batch}, and parametric ReLUs following previous literature~\cite{he2015delving}. 
We feed the input feature $\textbf{X}$ into this encoder and encode the $\textbf{X}$ into a series of hidden representations $\textbf{H}  = \{h_m^t\}$ for certain micro-crack on certain time steps respectively. After extracting the hidden representations $\textbf{H}  = \{h_m^t\}$,  we feed them into the recurrent transition layer (RTL),  which is built based on LSTM structure~\cite{LSTM}. 
We obtain the new hidden representation $\tilde{\textbf{H}}=\{\tilde{h}_{m,}^t\}$ in this RTL structure. Lastly, we continue to move the obtained $\tilde{\textbf{H}}$ to the expansive path (decoder) for outputting reconstructed fracture data $\textbf{Y}$. In particular, the expansive path consists of an upsampling of the feature map followed by a 2x2 convolution (“up-convolution”) that halves the number of feature channels, a concatenation with the correspondingly cropped feature map from the contracting path, and two 3x3 convolutions, each followed by a ReLU activation. The overall framework details are shown in Table.~\ref{Framework}.

The HOSSnet model outputs a reconstructed fracture data $\textbf{Y}$, the model is optimized to reduce the difference between obtained $\textbf{Y} = \{y_{m}^{t}\}$ and provided ground truth fracture data $\hat{\textbf{Y}} = \{\hat{y}_{m}^{t}\}$ for certain micro-crack $m$ on certain time step $t$. Such a difference is represented as a loss $\mathcal{L}_\text{MSE}(\textbf{Y},\hat{\textbf{Y}})$, which can be implemented as mean squared loss (MSE) that measure the difference between two sets of data. The loss function can be represented as

\begin{equation}
    \mathcal{L}_{\text{MSE}}(\textbf{Y}, \hat{\textbf{Y}}) = \frac{1}{|\textbf{Y}|} \sum_{\{(m,t)|y_{m}^{t}\in \textbf{Y}\}} (y_{m}^{t}-\hat{y}_{m}^{t})^2%,
    \label{eq:recon_loss}
\end{equation}
%where .... 

\subsection{Physics-Guided Regularization}
We further regularize our proposed model by leveraging the physical knowledge: optical flow~\cite{horn1981determining} and positive direction to optimize the reconstruction performance by several approaches described below.

\subsubsection{Optical Flow}
%\textcolor{red}{need help from Shihang}

Optical flow has been used as a method to describe the movement of an object between consecutive frames of the images~\cite{horn1981determining}, and we adopt it to represent the development of the fracture $\mathbf{Y}$. At each pixel location, the optical flow vector represents the motions of the pixel from one frame to another. Specifically, we consider each pixel in the fracture image $\mathbf{Y}(x,y,t)$  moving to $\mathbf{Y}(x+\Delta{x},y+\Delta{y},t+\Delta{t})$ after the time interval $\Delta{t}$. If the movements %\Delta{Y}$ and $\Delta{y}$ 
is small, $\mathbf{Y}(x+\Delta{x},y+\Delta{y},t+\Delta{t})$ can be expanded using a Tayler series:

\begin{equation}
\mathbf{Y}(x+\Delta{x},y+\Delta{y},t+\Delta{t})=
\mathbf{Y}(x,y,t)+\mathbf{Y}_{x}\Delta{x}+\mathbf{Y}_{y}\Delta{y}+\mathbf{Y}_{t}\Delta{t}+\ldots,
\label{eq:p_taylor}
\end{equation}
where $\mathbf{Y}_{x}$, $\mathbf{Y}_{y}$ and $\mathbf{Y}_{t}$ are partial derivative with respect to $x$, $y$ and $t$. %The pixel keeps the same after the movement and the higher order terms~(i.e.,``...'' in Eq.~(\ref{eq:p_taylor})) are ignored, it follows
If we consider the target pixel to be the same before and after the movement, we have $\mathbf{Y}(x+\Delta{x},y+\Delta{y},t+\Delta{t}) = \mathbf{Y}(x,y,t)$. Combining with Eq.~(\ref{eq:p_taylor}) by ignoring the higher order terms (i.e.,``...''), we have
\begin{equation}
\mathbf{Y}_{x}u+\mathbf{Y}_{y}v+\mathbf{Y}_t=0,
\label{eq:optical}
\end{equation}
where $u$ and $v$ are the $x$ and $y$ components of the motion vector $\mathbf{m} = (u,v)$, which is defined as
\begin{equation}
u=\frac{\Delta{x}}{\Delta{t}}, v=\frac{\Delta{y}}{\Delta{t}}.
\label{eq:motion}
\end{equation}
The motion vector $\mathbf{m} = (u,v)$ can be obtained by minimizing the objective function with a smooth regularization,
\begin{equation}
\zeta^\mathrm{op}=\iint\left(\left(\mathbf{Y}_{x}u+\mathbf{Y}_{y}v+\mathbf{Y}_t\right)^2+\lambda^2M^2\right)dxdy,
\label{eq:op_obj}
\end{equation}
where $\lambda$ is the weight for the smoothing regularization, and $M$ is the magnitude of the flow gradient given by
\begin{equation}
M({u},{v})=\sqrt{\|\nabla{u}\|^2+\|\nabla{v}\|^2}.
\label{eq:op_m}
\end{equation}

In this paper, we calculate the observed optical flow vector$~\mathbf{m_1}=(u_1,v_1)$ from the observed fractures $\hat{\mathbf{Y}}$ and the predicted optical flow vector $~\mathbf{m_2}=(u_2,v_2)$ from the predicted fractures $\mathbf{Y}$. Then we use the included angle between these two vectors as the regularization in the optical loss function $\mathcal{L}_\mathrm{op}$, which can be described as
\begin{equation}
\mathcal{L}_\mathrm{op}=\sum_{n}\left \|\arccos({r}_n (u_{1}, u_{2}, v_{1}, v_{2}))\right \|^2,
\label{eq:loss_op}
\end{equation}
where the operator $\arccos(\cdot)$ is the inverse of the cosine function. The function of ${r}_n (u_{1}, u_{2}, v_{1}, v_{2})$ is defined as 
\begin{equation}
{r}_n (u_{1}, u_{2}, v_{1}, v_{2})=\frac{u_{1}u_{2}+v_{1}v_{2}}{\sqrt{u_1^2+v_1^2}\sqrt{u_2^2+v_2^2}},
\label{eq:angle_n1}
\end{equation}
Thus, the overall loss function for the extrapolation network with optical flow regularization can be posed as
\begin{equation}
\mathcal{L}_\mathrm{MSE_{op}}=\mathcal{L}_\mathrm{MSE}+\alpha_{op}\,\mathcal{L}_\mathrm{op},
\label{eq:loss_e_reg}
\end{equation}
where the terms of $\mathcal{L}_\mathrm{MSE}$ and $\mathcal{L}_\mathrm{op}$ are provided in Eqs.~\eqref{eq:recon_loss} and \eqref{eq:loss_op}, respectively, and $\alpha_{op}$ is the weight for the regularization term.  The regularization term would constrain the direction of the fracture development.

\begin{table}[!t]
\small
\newcommand{\tabincell}[2]{\begin{tabular}{@{}#1@{}}#2\end{tabular}}
\centering
\caption{The detailed parameters' information of overall HOSSnet Framework including encoder(upper), recurrent transition model(middle), and decoder(lower).}
%\vspace{.1in}
\begin{tabular}{|p{5.0cm}|p{2.5cm}|p{2.5cm}|p{2.0cm}|}
\hline
\textbf{Layers} & Filter Size & State size   \\ \hline 
Convolutional layer & (3, 3)& 64\\
Residual Block& (3, 3)& 64\\
Residual Block& (3, 3)& 64\\
Residual Block& (3, 3)& 64\\
Maxpooling& None& None\\
Convolutional layer & (3, 3)& 64\\
Residual Block& (3, 3)& 64\\
Residual Block& (3, 3)& 64\\
Residual Block& (3, 3)& 64\\
Convolutional layer & (3, 3)& 64\\
\hline
LSTM Layer &None & 64\\
\hline
Convolutional layer & (3, 3)& 64\\
Upsampling & None&None \\
Add & None&None \\
Residual Block& (3, 3)& 64\\
Residual Block& (3, 3)& 64\\
Residual Block& (3, 3)& 64\\
Convolutional layer & (3, 3)& 64\\
Fully-Connection layer &None & None\\
\hline
\end{tabular}
\label{Framework}
%\vspace{-.1in}
\end{table}

\subsubsection{Positive Direction}

Ideally, the fracture development should always be positive with time changes. Without the awareness of such development patterns, the neural network model can still output negative changes. In order to force the fracture to develop in a positive direction, we have eliminated the negative change points during the prediction stage and set these points the same as the previous time step.%we have set a much higher weight for the negative changes in the loss function, which we change the equation~\ref{eq:recon_loss} as:
%\begin{equation}
%    \mathcal{L}_{\text{MSE}}(\textbf{Y}, \hat{\textbf{Y}}) = \frac{1}{|\textbf{Y}|} \sum_{\{(m,t)|y_{m}^{t}\in \textbf{Y}\}} \alpha_{pos}(y_{m}^{t}-\hat{y}_{m}^{t})^2.
 %   \label{eq:postive_loss}
%\end{equation}
%where $\alpha_{pos}$ is a large value if $\hat{y}_{m}^{t}<{y}_{m}^{t-1}$ and a small value if $\hat{y}_{m}^{t}>{y}_{m}^{t-1}$. 

\subsection{Machine Learning Optimization Approaches}

\subsubsection{Perceptual Loss}
In order to further improve the reconstruction performance, we not only use the content loss based on MSE but also introduce the perceptual loss calculated based on the feature map of the VGG network~\cite{simonyan2014very}, which works by summing all the squared errors between all the pixels and taking the mean. This is in contrast to a pixel-wise loss function(e.g., MSE) which sums all the absolute errors between pixels. This perceptual loss $ \mathcal{L}_{\text{Perceptual}}$ can be represented as:

\begin{equation}
    \mathcal{L}_{\text{Perceptual}}(\textbf{E}, \hat{\textbf{E}}) = \frac{1}{|\textbf{E}|} \sum_{\{(m,t)|e_{m}^{t}\in \textbf{E}\}} (v_{m}^{t}-\hat{e}_{m}^{t})^2,
    \label{eq:recon_loss}
\end{equation}
where $\textbf{E}$ and $\hat{\textbf{E}}$ represent the pixel features extracted from VGG network for $\textbf{Y}$ and $\hat{\textbf{Y}}$ respectively. Hence, the loss function $\mathcal{L}_\text{recon}$ can be replaced by
\begin{equation}
    \mathcal{L}_{\text{recon}}(\textbf{Y}, \hat{\textbf{Y}}) = \mathcal{L}_{\text{MSE}}(\textbf{Y}, \hat{\textbf{Y}}) + \alpha_{Perc}\mathcal{L}_{\text{Perceptual}}(\textbf{Y}, \hat{\textbf{Y}}),
    \label{eq:percep_loss}
\end{equation}
where $\alpha_{Perc}$ is the weight of perceptual loss. Overall, after adding all the regularization and physical information,  the loss function for the optimization will be

\begin{equation}
    \mathcal{L}_{\text{all}}(\textbf{Y}, \hat{\textbf{Y}}) = \mathcal{L}_{\text{MSE}}(\textbf{Y}, \hat{\textbf{Y}}) 
    +\alpha_{Perc}\mathcal{L}_{\text{Perceptual}}(\textbf{Y}, \hat{\textbf{Y}})
    + \alpha_{op}\,\mathcal{L}_\mathrm{op}(\textbf{Y}, \hat{\textbf{Y}}).
    \label{eq:all_loss0}
\end{equation}

\subsubsection{Sub-region}
In addition, when we utilize the extrapolation and the interpolation task, we know the location of the fracture from the known data. Instead of training with the whole image, %which mostly has no changes, 
we train the network with the sub-region of the images $\textbf{Y}_{sub}$ containing the fracture dynamic. This is because the most of region of the whole image has no change, making the built HOSSnet model less sensitive to the sub-region containing fracture dynamic. Hence, the loss function after adding this optimization will be replaced as:

\begin{equation}
    \mathcal{L}_{\text{all}}(\textbf{Y}_{sub}, \hat{\textbf{Y}}_{sub}) = \mathcal{L}_{\text{MSE}}(\textbf{Y}_{sub}, \hat{\textbf{Y}}_{sub}) 
    +\alpha_{Perc}\mathcal{L}_{\text{Perceptual}}(\textbf{Y}_{sub}, \hat{\textbf{Y}}_{sub})
    + \alpha_{op}\,\mathcal{L}_\mathrm{op}(\textbf{Y}_{sub}, \hat{\textbf{Y}}_{sub}).
    \label{eq:all_loss}
\end{equation}

\section{Experiment}
This section will first describe the HOSS dataset and the experiment settings. Then we evaluate the performance of our proposed methods for fracture reconstruction.

\subsection{Experiment Setting}

We consider two experiments: Extrapolation and Interpolation. The former is designed to verify the ability of the proposed method to reconstruct micro-crack fractures in the future state. We have conducted experiments over the sample, in which the test is performed on different samples from the training set. In addition, The experiments over time are performed, wherein predictions were made on the same sample, but across different time periods from the training set. The latter is designed to verify the ability of the proposed method to reconstruct micro-crack fracture in the intermediate time period.

\begin{figure} [!h]
\centering
% \vspace{-.1in}
%\vspace{-.1in}
\includegraphics[width=0.3\linewidth]{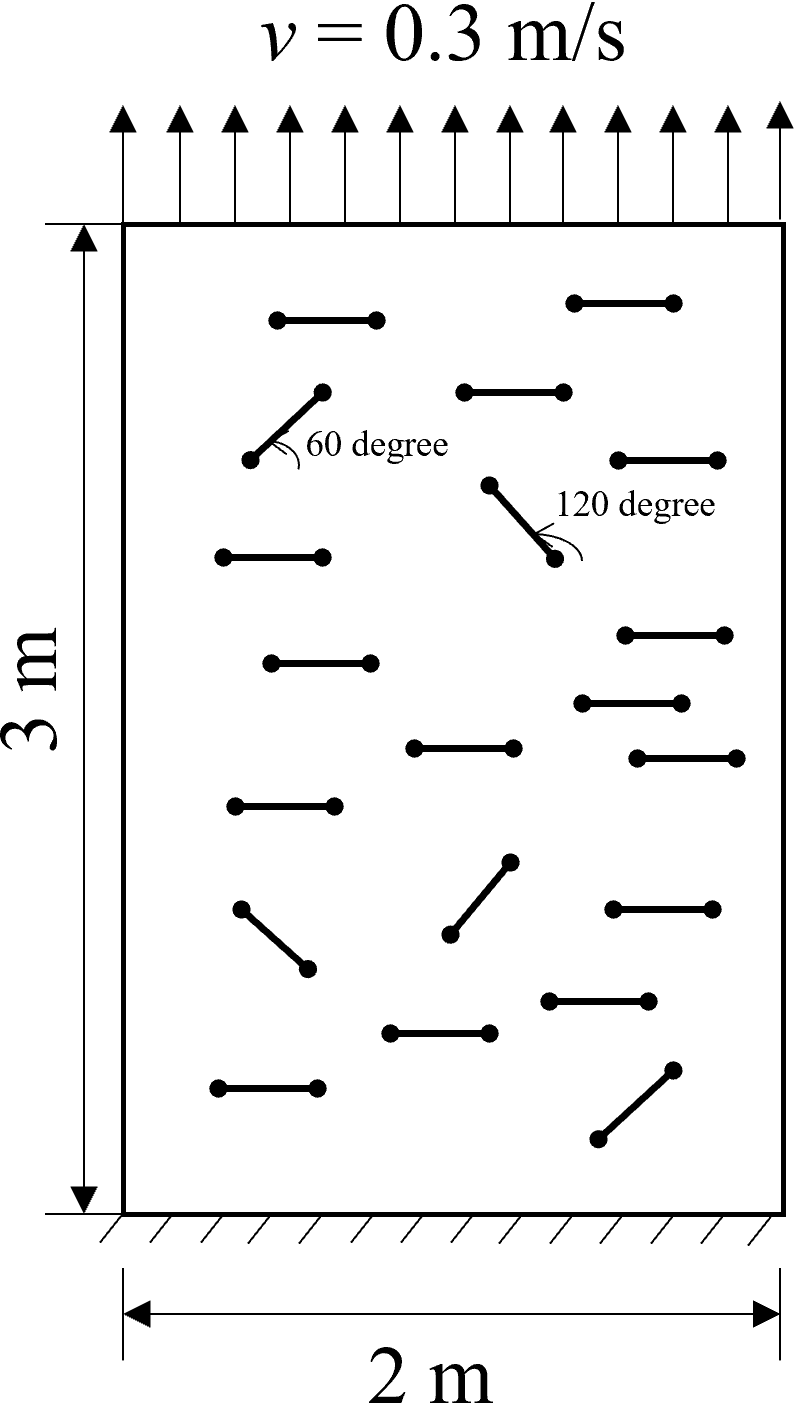}
%\vspace{-.1in}
\caption{Model setup for the 2D uni-axial tensile failure problem with 20 pre-existing cracks.  } %\textcolor{red}{need to redraw}}
%\vspace{-.1in}
\label{fig:crack}
\end{figure}

\subsubsection{Dataset}

The high-fidelity HOSS simulator has been applied to simulate the fracture growth and material failure of a concrete sample under uniaxial tensile loading conditions. Then, the generated simulation data are converted to images by ParaView for the training of HossNet. The problem is considered a 2D problem, and the model setup is shown in Figure~\ref{fig:crack}. The domain is a rectangular sample $\text{2m}$ wide and $\text{3m}$ high. The bottom edge of the sample is fixed. The top edge of the sample moves upwards with a constant velocity of $\text{0.3m/s}$, which results in a strain rate of $\text{0.1s}^{-1}$. The material is assumed to be elastically isotropic for all cases with the elastic material properties: $\text{2500kg/}{\text{m}^3}$ for density, 22.6 GPa for Young's modulus, and 0.242 for Poisson's ratio.

\subsubsection{Evaluation Metrics}

We evaluate the performance of micro-reconstruction using three different metrics, root mean squared error (RMSE), structural similarity index measure (SSIM)~\cite{wang2004image}, and weighted fracture error (WFE). We use RMSE to measure the difference (error) between reconstructed data and target fracture data. The lower value of RMSE indicates better reconstruction performance. SSIM is used to appraise the similarity between reconstructed data and target fracture on three aspects, luminance, contrast, and overall structure. The higher value of SSIM indicates better reconstruction performance. Lastly, we also create a new evaluation metric: weighted local fracture error (WFE),  to measure the difference by using mean squared error, given the higher weight of the local region with dynamic change of micro-crack fracture. This evaluation metric can be represented as:

\begin{equation}
\small
\text{WFE} = 10\text{RMSE}(\textbf{Y}_{dym} - \hat{\textbf{Y}}_{dym}) + \text{RMSE}(\textbf{Y}_{fix} - \hat{\textbf{Y}}_{fix}),
\end{equation}
where $\textbf{Y}_{dym}$ and $\textbf{Y}_{fix}$ denote the dynamic region and fixed region, respectively.
The local crack prediction performance is more effectively represented by the WFE metric as compared to RMSE due to its emphasis on the local region. Similar to RMSE, the lower value of WFE indicates better reconstruction performance as well.

\subsubsection{Baselines}

We compare the performance of the proposed HOSSnet method against several existing methods that have been widely used for data reconstruction. Specifically, we implement the HRU, and CNN-LSTM as baselines. To better verify the effectiveness of each component in our proposed method, we further compared HOSSnet with its variants: HOSSnet-F, which removes the perception loss from the complete model and only includes physical regularization in the proposed method. Additionally, by comparing HOSSnet-F and CNN-LSTM, we show improvement by incorporating physical regularization. We can further verify the effectiveness of the perception loss by comparing HOSSnet-F and HOSSnet.

\subsubsection{Experimental Design}
We evaluate the performance of our proposed method in two different scenarios.

%\textcolor{red}{Shihang: can you add your design of interpolating experiment in this part?}

First, we conduct the extrapolation experiment, which includes over-time and over-sample tests, to study how the proposed HOSSnet method helps simulate fracture change in both over-time and over-sample situations. Specifically, in the over-time test, we select 6 different cracks in our test. We use the 5 complete crack samples and the first 150 time steps of the sixth crack sample as training and reconstruct the crack data in the remaining 150-time steps. In the over-sample test, we also use 6 different crack samples, each sample containing 300-time steps crack data. We use the 5 different crack samples as training and reconstruct the complete sequence of the remaining crack sample. Both over-time and over-sample tests are challenging tasks since micro-crack fracture is changing over time following complex non-linear patterns (driven by a partial differential equation). Hence, in the over-time test, the model trained from available data may not be able to generalize to future data that look very different from training data. Furthermore, it is more difficult to achieve the complete sequence of data reconstruction in the over-sample test because the model cannot capture any pattern of the testing data. 

Second, we consider the case in which the intermediate time period of data is missing. For example, the fracture data is available at the beginning and end time periods but missing in the intermediate time steps. We can use the model trained using available data (beginning and ending) to reconstruct the fracture data in the missing time steps. 

We refer to this test as an interpolation experiment. In this test, we use the single crack sample in our test, containing 300 available time steps. In the first case, we divide the data into 3 parts and each part has 100-time steps of data.  We use the first and last 100 time steps as training and use the remaining intermediate 100-time steps for testing. In the other case, we only use $10\%$ time steps as the training set, which we include the dataset into the training set every 10-time steps, the remaining are the test set. 

\subsubsection{Training Settings}

Data normalization is performed on both training and testing datasets, to normalize input variables to the range [0, 1]. Then, the model is trained by ADAM optimizer~\cite{kingma2014adam_arxiv} The initial learning rate is set to 0.0005 and iterations are 500 epochs. We use Tensorflow 2.6 and Keras to implement our models with V100 GPU.

\begin{table}
\centering
\caption{Average reconstruction performance (measured by RMSE,  SSIM, and WFE) of first 50th time steps in Cauchy $\rightarrow$ Fracture's experiment on extrapolating over the sample (upper),  extrapolating over time (middle), interpolating (lower) respectively.}
%\vspace{.1in}
\label{cauchy}
\begin{tabular}{|c|c|c|c|c|} 
\hline
Experiment                     & Method    & RMSE$\downarrow$ & SSIM$\uparrow$ & WFE$\downarrow$  \\ 
\hline
\multirow{4}{*}{Over Sample}   & HRU     &     0.057              &      0.965          &      0.091            \\
                               & CNN-LSTM  & 0.046           & 0.971          & 0.083            \\
                               & HOSSnet-F & 0.038           & 0.979          & 0.077            \\
                               & HOSSnet   &     \textbf{0.028}              &      \textbf{0.982}           &        \textbf{0.060}           \\ 
\hline
\multirow{4}{*}{Over Time}     & HRU       &    0.023              &     0.985           &     0.061             \\
                               & CNN-LSTM  &    0.018              &     0.985           &     0.052            \\
                               & HOSSnet-F & 0.011                &       0.991         &       0.034       \\
                               & HOSSnet   &    \textbf{0.009}               &    \textbf{0.995}             &         \textbf{0.024}          \\ 
\hline
\multirow{4}{*}{Interpolation} & HRU       &   0.025               &      0.985          &       0.057           \\
                               & CNN-LSTM  &       0.020           &  0.987              &    0.046              \\
                               & HOSSnet-F &    0.018              &    0.988            &      0.049            \\
                               & HOSSnet   &     \textbf{0.014}              &      \textbf{0.991}           &    \textbf{0.036}               \\
\hline
\end{tabular}
%\vspace{-.1in}
\end{table}

\subsection{Cauchy 1, 12, 2 $\rightarrow$ Fracture}
This section shows the experimental results of Cauchy features $\rightarrow$ Fracture, which also includes extrapolating over the sample, over time, and interpolating experiments.

\subsubsection{Extrapolation Over Sample}

\begin{figure*} [ht]
\centering
%\raggedleft
\subfigure[RMSE.]{ \label{fig:a}{}
\includegraphics[width=0.48\linewidth]{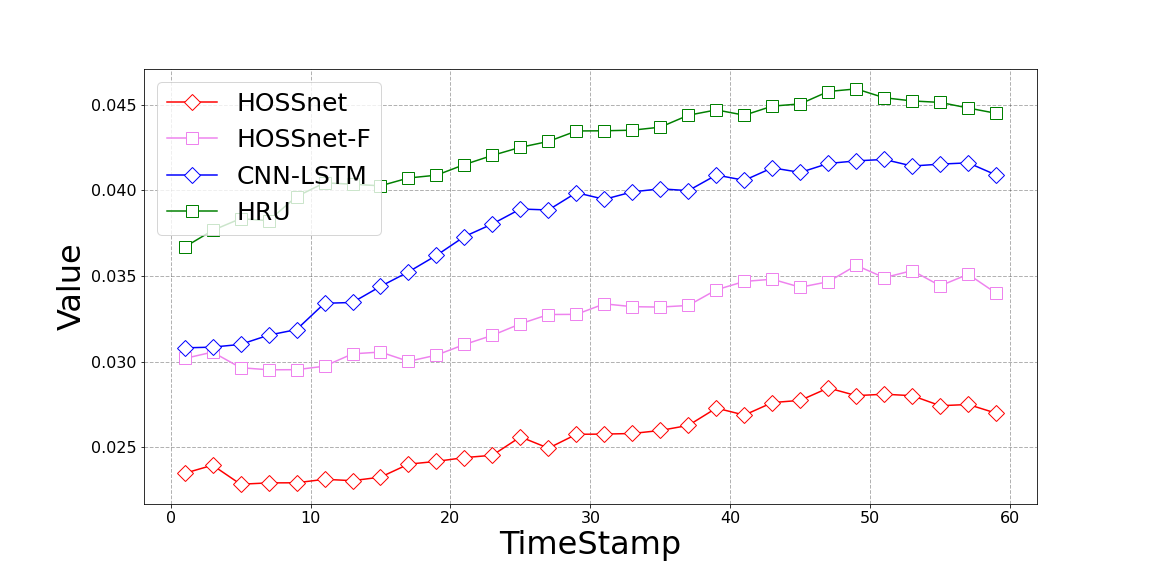}
}%\hspace{-0.1in}
\subfigure[LFE.]{ \label{fig:b}{}
\includegraphics[width=0.48\linewidth]{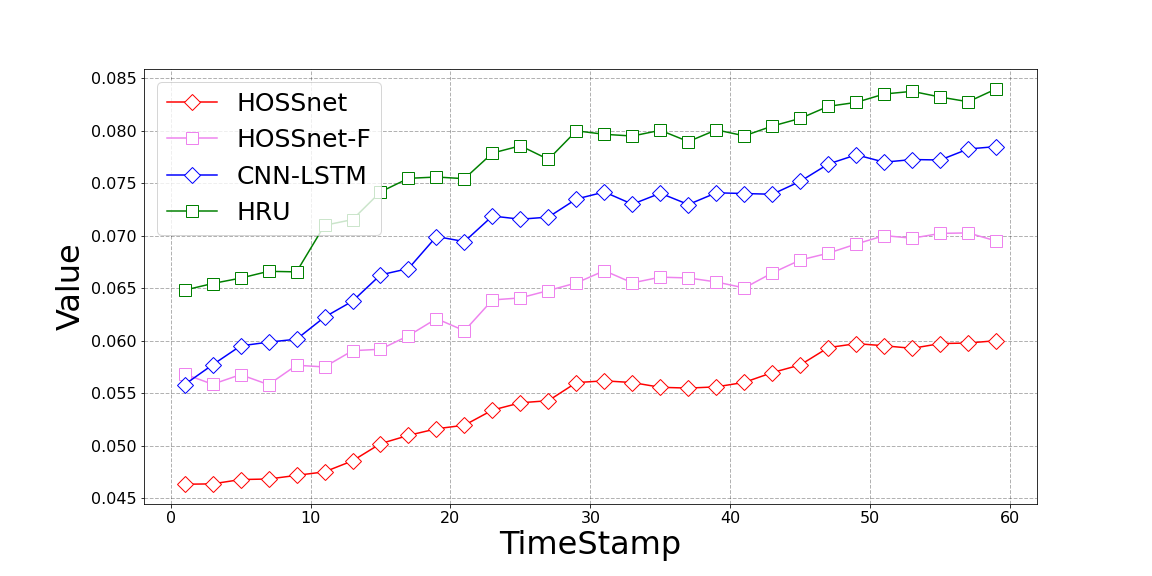}
}%\hspace{-0.1in}
\vspace{-.1in}
\caption{Change of RMSE and WFE values produced by different models from the 1st to the 60th time (interval is 2 time steps) step in the extrapolating over sample experiment.}
\label{fig:oversample}
\end{figure*}

%\begin{comment} 
%\begin{figure*}[h]
%\begin{center}
%% \raggedleft
%\includegraphics[width = 0.9\linewidth]%{figures/Shengyu/Cross_pred/Cross_pred.png}
%\end{center}
%%\vspace{-.2in}
%\caption{Reconstructed fracture images by HOSSnet Model~(top) and real fracture damages (bottom) from the 1st to the 50th time steps (interval is 10 time steps) in the Cauchy$\rightarrow$Fracture over sample scenario.}
%\label{fig:cross_pred}
%\vspace{-.1in}
%\end{figure*}
%\end{comment}

\begin{figure*} [ht]
\centering
%\raggedleft
\subfigure[Predicted Fracture.]{ \label{fig:a}{}
\includegraphics[width=0.9\linewidth]{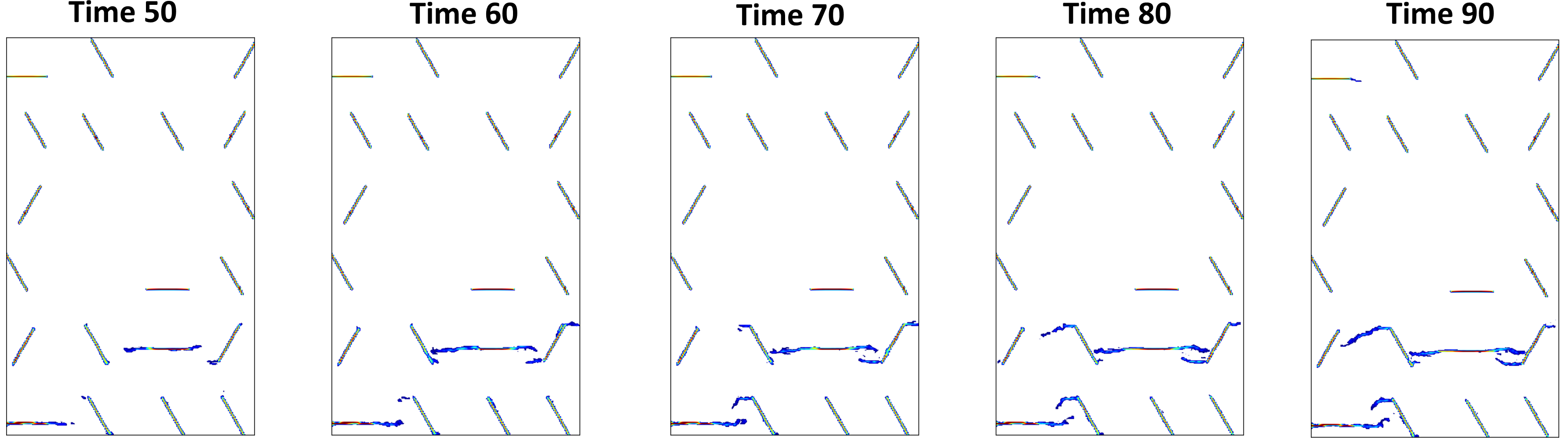}
}\vspace{-.05in}
\subfigure[Ground Truth.]{ \label{fig:b}{}
\includegraphics[width=0.9\linewidth]{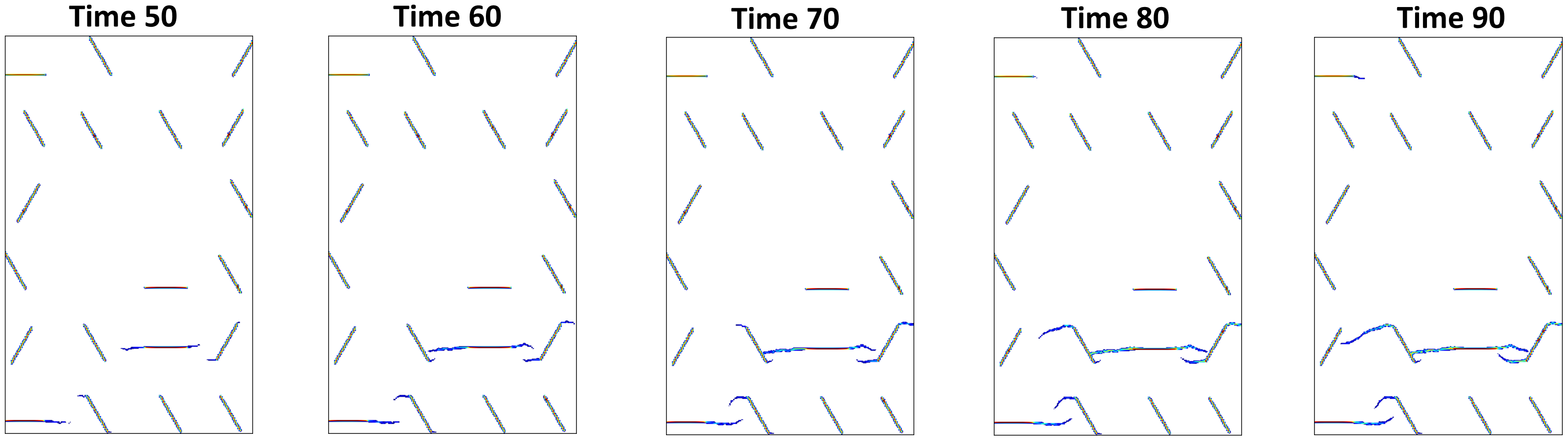}
}\vspace{-.05in}
\subfigure[Difference.]{ \label{fig:c}{}
\includegraphics[width=0.9\linewidth]{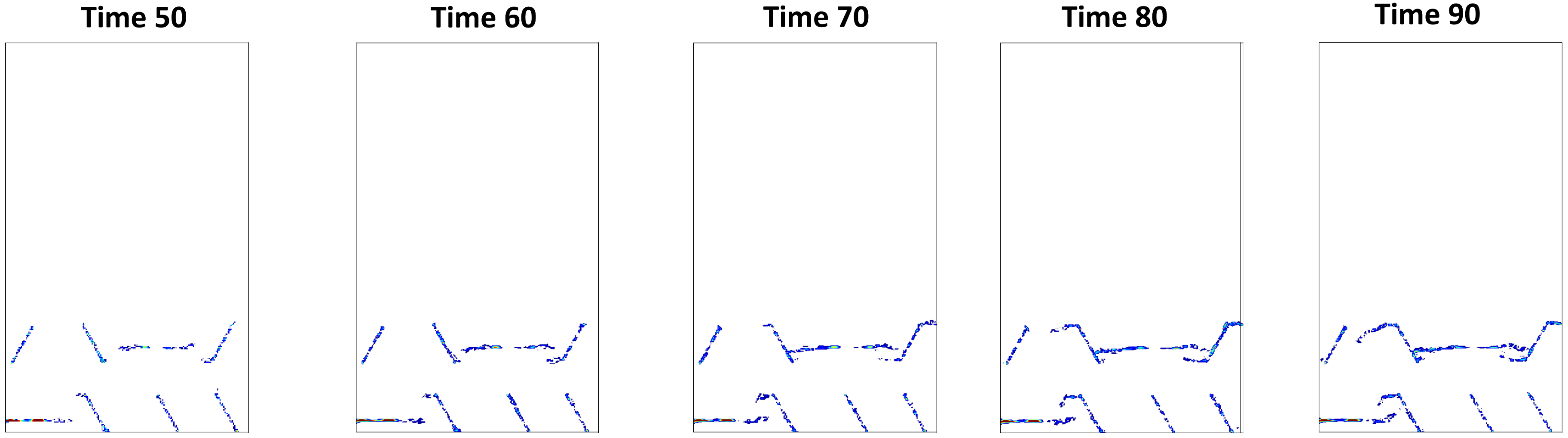}
}%\hspace{-0.1in}
%\vspace{-.1in}
\caption{Reconstructed fracture images by HOSSnet Model~(top), real fracture images (middle), and difference  (bottom) between reconstructed fracture and real fracture from the 1st to the 50th time steps (interval is 10 time steps) in the Cauchy$\rightarrow$Fracture over sample scenario.}
\label{fig:cross_pred}
\end{figure*}

\textbf{Quantitative Results.} The upper section of Table.~\ref{cauchy}, shows quantitative comparisons amongst all the methods, including average RMSE, SSIM, and WFE in the first 50-time steps of the testing phase. When comparing our proposed HOSSnet method with baseline methods. our proposed HOSSnet performs the best in three evaluation ways: obtain the lowest RMSE and WFE values, and the highest SSIM values. According to this table, the proposed method HOSSnet, in general, outperforms other baselines in terms of three different evaluation metrics. By comparing HRU and CNN-LSTM, we show the improvement by using the RTL structure. Furthermore, the comparison amongst CNN-LSTM, HOSSnet-F, and HOSSnet shows the effectiveness of incorporating the physics-guided regularization method (optical flow and perpetual loss) of the proposed method. In particular, the incorporation of perpetual loss brings the most significant performance improvement in terms of RMSE, SSIM, and WFE values.

\textbf{Temporal Analysis}
In the temporal analysis, we show the change in performance as we reconstruct fracture data over 60-time steps after the training data. We show the performance of long-term prediction in terms of RMSE and WFE in  Figures.~\ref{fig:oversample}. Several observations are highlighted: (1) With larger time intervals between training data and prediction data, the performance becomes worse. In general, our proposed HOSSnet method outperforms other baselines. (2) We can obtain a similar conclusion by comparing HRU and CNN-LSTM. The proposed RTL component can bring significant improvement in long-term propagation. (3) We also can justify the effectiveness of the proposed physics-guided regularization method and extra perpetual loss by comparison amongst CNN-LSTM, HOSSnet-F, and HOSSnet.

\textbf{Visual Results.}
 Figures.~\ref{fig:cross_pred} shows the reconstructed fracture at multiple time steps (50th, 60th, 70th, 80th, and 90th-time steps) after the training period. For each time step, we show the images from ground truth and the HOSSnet method. In the first few time steps, we can observe that our proposed HOSSnet can obtain ideal visual results. The performance is very similar to the ground truth image. After several time steps (e.g., 70th-time step), we can find the reconstruction performance is become slightly worse than the ground truth image but still can effectively fine-level captures textures and patterns, and obtain reasonable performance, only containing a slight blur in the sub-region of the dynamic of fracture. It also proves the effectiveness of the proposed HOSSnet method in long-term reconstruction.

\subsubsection{Extrapolation Over Time}

\textbf{Quantitative Results.} In this experiment, we compare the same set of existing methods as in the previous Extrapolation Over Sample experiment. In the middle part section of Table~\ref{cauchy}, we show the quantitative results of the first 50-time steps in the testing set. We can observe similar results that our proposed HOSSnet outperforms other methods by a considerable margin. We still can notice a significant improvement by adding each component: RTL, Physics-Guided Regularization, and perpetual loss, but the degree of improvement is not as high as in the over-sample experiments. This is because we also use the first 150-time steps of testing crack data as training except for 5 complete sequences of crack data. This is very helpful for the ML model to capture the initial complex pattern and dynamic transformation of the testing data. In such a case, a simple HRU model can get a relatively reasonable result. Therefore, the improvement brought by adding other components later will not be as obvious as in the previous over-sample experiment. 

We also show the details of temporal analysis and visual results for this extrapolating over-time experiment in the supplementary file. We almost can obtain a similar conclusion as the experiment of extrapolating over the sample.  Our proposed method can achieve promising reconstruction performance in long-term propagation.

\subsubsection{Interpolation}

\textbf{Quantitative Results.}
In this single crack experiment, we compare the same set of existing methods as in the previous two extrapolating experiments. In the lower part section of Table~\ref{cauchy}, we show the quantitative results including average RMSE, SSIM, and WFE in the first 50 time steps of the testing set.  From the table, we also obtain the same conclusion that our proposed HOSSnet model can outperform other baselines and each new component can bring significant improvement in terms of these three evaluation metrics. In addition, we can observe that the overall performance from the interpolation experiment is slightly worse than the extrapolation over-time experiment. This is because we only conduct interpolating tests in a single-crack experiment and use the beginning and end of crack data as training. It causes ML models to have difficulty capturing the complete dynamics of the crack without adding additional complete sequences of crack data (similar to extrapolating experiments' settings).

We also show the details of temporal analysis and visual results for this interpolation experiment in the supplementary file. We almost can obtain a similar conclusion as quantitative analysis. Our proposed method can achieve promising reconstruction performance in this interpolation condition.

\begin{figure*}[htbp]
\begin{center}
% \raggedleft
\includegraphics[width = 1.0\linewidth]{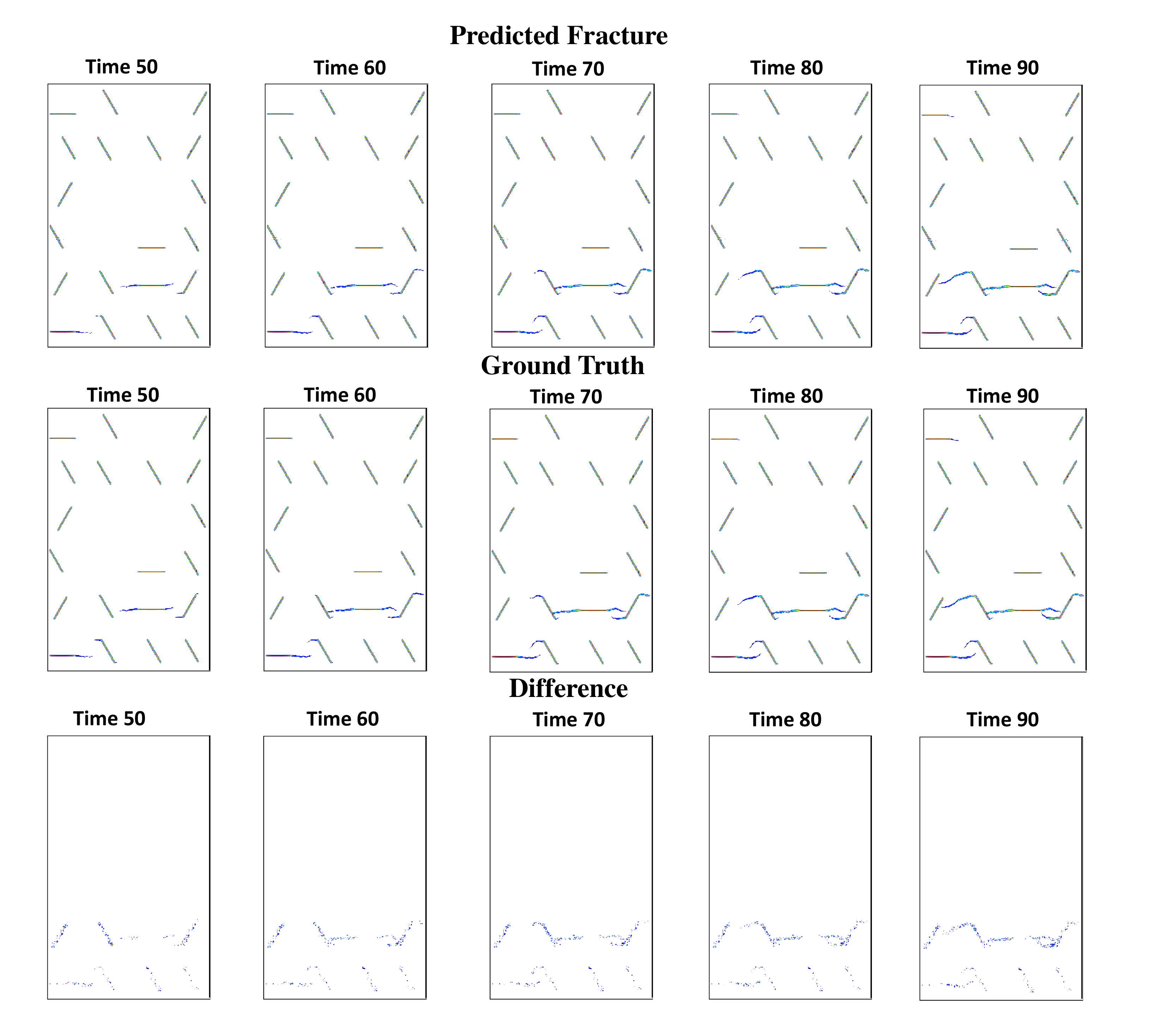}
\end{center}
\vspace{-.3in}
\caption{Reconstructed fracture images by HOSSnet Model and real fracture images from the 1st to the 50th
time steps (interval is 10-time steps) in the Fracture$\rightarrow$Fracture over sample scenario.}
\label{fig:crack_predict}
\vspace{.1in}
\end{figure*}

\subsection{Fracture $\rightarrow$ Fracture }

This section represents the experimental results with Fracture $\rightarrow$ Fracture, which includes extrapolation over the sample and over time. The setting of these experiments is the same with Cauchy features $\rightarrow$ Fracture except the inputs are fracture damage.

\subsubsection{Extrapolation Over Sample}

Table~\ref{perf1} gives the RMSE, SSIM, and WFE of all the networks. From the results, we can see the improvement of the network with different regularization terms. When considering solely the implementation of physical regularization, the HOSSnet-F model does not exhibit a notable improvement in performance as compared to the CNN-LSTM model. With all the regularization terms, our proposed HOSSnet method still yields the lowest RMSE and WFE and the highest SSIM. The examples of the predicted fracture from HOSSNet at different time steps are shown in Figure~\ref{fig:crack_predict}. Compared to the scenario that uses Cauchy stress as the input, the network performs better when the input is the fracture because both input and output have the same properties. However, the damage is harder to obtain in the survey which makes this scenario harder to implement in the real world.

\begin{table}[ht]
\vspace{.1in}
\centering
\caption{Reconstruction performance  (measured by RMSE,  SSIM, and WFE) of Fracture $\rightarrow$ Fracture's experiment on the extrapolating over the sample (upper),  extrapolating over time tests (middle), interpolating (lower) respectively.}
%\vspace{.1in}
\label{perf1}
\begin{tabular}{|c|c|c|c|c|} 
\hline
Experiment                     & Method    & RMSE$\downarrow$ & SSIM$\uparrow$ & WFE$\downarrow$  \\ 
\hline
\multirow{4}{*}{Over Sample}   & HRU       &0.029                  &0.976                & 0.124                 \\
                               & CNN-LSTM  & 0.026            & 0.985          & 0.058            \\
                               & HOSSnet-F &0.023             &    0.986       & 0.074            \\
                               & HOSSnet   &  \textbf{0.013}                 &     \textbf{0.993}            &   \textbf{0.045}                \\ 
\hline
\multirow{4}{*}{Over Time}     & HRU       &0.010                  &   0.992             &    0.051              \\
                               & CNN-LSTM  &    0.008              &     0.996           &     0.051             \\
                               & HOSSnet-F & 0.007                &       0.998         &       0.031       \\
                               & HOSSnet   &  \textbf{0.006}                 &     \textbf{0.999}            &        \textbf{0.030}           \\ 
\hline
\end{tabular}
%\vspace{-.1in}
\end{table}

% For our proposed HOSSnet, the SSIM can achieve a high SSIM as 

\subsubsection{Extrapolation Over Time}

According to the results presented in Table~\ref{perf1}, the HOSSNet demonstrates the most superior performance. This is attributed to the fact that both the training and testing phases are carried out at distinct temporal intervals within the same crack. Consequently, the HOSSNet prediction exhibits the highest SSIM of 0.999 compared to other scenarios. Nonetheless, it is worth noting that a distinct network must be trained for each crack to be applied to diverse fractures, leading to increased computational expenses for this scenario.

\section{Conclusion}
In this paper, we develop a new data-driven method HOSSnet that integrates a series of physics regulations: optical flow, and positive direction to achieve fracture reconstruction in spatial-temporal. We also introduce extra machine learning optimization approaches to further improve the reconstruction. Moreover, we come up with two different experiment settings: Cauchy $\rightarrow$ Fracture and Fracture $\rightarrow$ Fracture to show the robustness and applicability of our proposed HOSSnet model. Our experiments demonstrate the effectiveness of these proposed methods in improving fracture reconstruction in extrapolating and interpolating tests. Furthermore,  we also will propose new extensions by incorporating the underlying physical relationships (e.g. underlying partial differential equation(PDE) into the proposed model to further improve the reconstruction performance and evaluate our proposed HOSSnet method in larger micro-crack regions.

\section*{Acknowledgments}
This work was funded by the Los Alamos National Laboratory~(LANL) - Laboratory Directed Research and Development program under project number 20210542MFR. %\yl{Other funding sources?}

%%Vancouver style references.
\bibliographystyle{model1-num-names}
\bibliography{refs}

\end{document}